\documentclass[twocolumn]{aa}
\usepackage{graphicx}
\usepackage{txfonts}
\usepackage{natbib}
\bibpunct{(}{)}{;}{a}{}{,}
\renewcommand{\object}[1]{#1}
\begin{document}
   \title{Diagnostics of active galaxies}
   \subtitle{I. Modeling the infrared properties of dusty cores of starburst galaxies}
   \titlerunning{Modeling the infrared properties of dusty cores of starburst galaxies}
   \author{A.F. Loenen\inst{1,2} \and  W.A. Baan\inst{2} \and M. Spaans\inst{1}
          }
   \offprints{A.F. Loenen}
   \institute{Kapteyn Astronomical Institute, P.O. Box 800, 9700 AV  Groningen, the Netherlands \\
             \email{loenen@astro.rug.nl}\\
             \email{spaans@astro.rug.nl}
             \and
             ASTRON, P.O. Box 2, 7990 AA Dwingeloo, the Netherlands \\
         \email{baan@astron.nl}
            }

   \date{Received 08/06/2006 / Accepted 27/07/2006}

   \abstract{}{Despite extensive observations over the last decades,
        the central questions regarding the power source of the large
        IR luminosity of Ultra Luminous Infra Red Galaxies (ULIRGs),
        and their evolution, are still not fully answered. In this
        paper we will focus on massive star formation as a central
        engine and present an evolutionary model for these
        dust-enshrouded star formation regions.}{An evolutionary model
        was created using existing star formation and radiative
        transfer codes (STARBURST99, RADMC and RADICAL) as building
        blocks. The results of the simulations are compared to data
        from two IRAS catalogs.}  {From the simulations it is found
        that the dust surrounding the starburst region is made up from
        two components. There is a low optical depth ($\tau=0.1$,
        which corresponds to 0.1 \% of the total dust mass), hot
        (T$\sim$400K) non-grey component close to the starburst (scale
        size 10pc) and a large scale, colder grey component (100pc,
        75K) with a much larger column ($\tau=10$).  The simulations
        also show that starburst galaxies can be powered by massive
        star formation. The parameters for this star forming region
        are difficult to determine, since the IR  continuum
        luminosity is only sensitive to the total UV input. Therefore,
        there is a degeneracy between the total starburst mass and the
        initial mass function (IMF) slope. A less massive star
        formation with a shallower IMF will produce the same amount of
        OB stars and therefore the same amount of irradiating UV
        flux. Assuming the stars are formed according to a Salpeter
        IMF ($\Psi(M) \propto M^{-2.35}$), the star formation region
        should produce $10^9$~M$_{\sun}$ ~of stars (either in one
        instantaneous burst, or in a continuous process) in order to
        produce enough IR radiation.} {Our models confirm that massive
        star formation is a valid power source for ULIRGs.  In
        order to remove degeneracies and further determine the
        parameters of the physical environment also IR spectral
        features and molecular emissions need to be included.}

   \maketitle
%

\section{Introduction}
\label{sec:introduction}
In 1983 the Infra-Red Astronomical Satellite (IRAS) surveyed 96\% of
the sky in four broad-band filters at 12$\mu$m, 25$\mu$m, 60$\mu$m,
and 100$\mu$m. IRAS detected infrared (IR) emission from about 25,000
galaxies, primarily from spirals, but also from quasars (QSOs),
Seyfert galaxies and early type galaxies. Among these galaxies IRAS
discovered a new class of galaxies that radiate most of their energy
in the infrared. The most luminous of these infrared galaxies, the
(ultra-)luminous infrared galaxies [(U)LIRGs], have QSO-like
luminosities of L $\ge 10^{11}$ L$_{\sun}$ ~(LIRGs) or even L $\ge 10^{12}$
L$_{\sun}$ ~(ULIRGs) \citep{2000ARA&A..38..761G}.

Despite extensive observations over the last decades, the central
questions regarding the power source of the large IR luminosity of
ULIRGs, and their evolution, are still not fully answered.
\cite{1988ApJ...325...74S} proposed that most ULIRGs are powered by
dust-enshrouded QSOs in the late phases of a merger. The final state
of such a merger would be a large elliptical galaxy with a massive
quiescent black hole at its center \citep{1992ApJ...390L..53K}. A
significant fraction of the ULIRG population seems to confirm this
assumption, since they exhibit nuclear optical emission line spectra
similar to those of Seyfert galaxies \citep{1988ApJ...325...74S}. Some
also contain compact central radio sources and highly absorbed, hard
X-ray sources, all indicative of an active nucleus (AGN).

On the other hand, the (Far)IR, mm, and radio characteristics of
ULIRGs are similar to those of starburst galaxies. A centrally
condensed burst of star formation activity (called a starburst,
hereafter denoted as SB), for instance fueled by gas driven into
the center of the potential well of a pair of interacting galaxies
by a bar instability, provides an equally plausible power source
\citep[e.g.][]{2002PhR...369..111B}.  Observational evidence for
the starburst nature of ULIRGs was found with the detection of a
large number of compact radio hypernovae in each of the two nuclei
of \object{Arp220} by \cite{1998ApJ...493L..17S}.

\begin{figure}
  \centering
  \includegraphics[angle=0,width=\hsize]{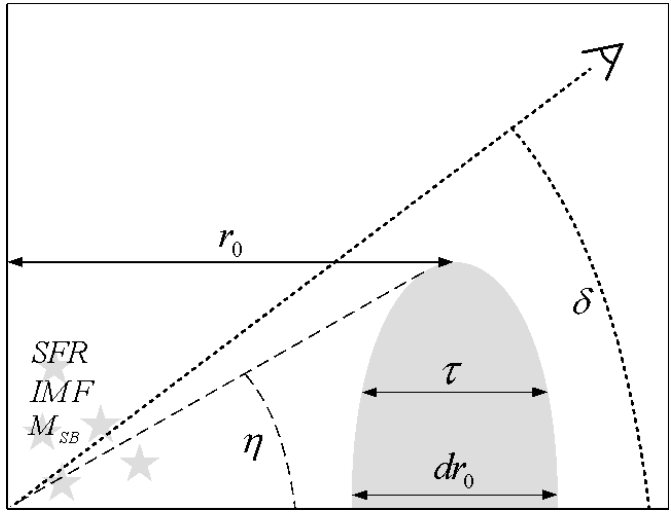}
  \caption{Schematic overview of the ``physical'' environment created
  in the simulations. In the center star formation is going on, which
  irradiates the dusty surroundings (here represented as a toroidal
  shape). This dust then re-radiates its given energy in the IR
  regime. Also the parameter space is shown. In the starburst, the
  total stellar mass ${\rm M_{SB}}$, the star formation rate (SFR) and
  the IMF-slope $\alpha$ are varied. In the dust the geometry is
  varied by changing the closing angle $\eta$. Also the dust column density
  (expressed as the optical depth $\tau$) is varied. A last variable
  is the observational inclination $\delta$. Also some parameters are
  shown, which initially were not varied: the radius $r_0$ and the radial scale
  size $dr_0$. More information on the parameter space can be found in
  Sect. \ref{sec:simulations}.}
\label{fig:setup}
\end{figure}

Since both options are observationally supported, it is natural to
think that there might be an evolutionary relation between the
two. Several authors have suggested such schemes. In
\citeyear{1988ApJ...330..743B}, \citeauthor{1988ApJ...330..743B}
explained the evolution of the FIR properties of active nuclei
with a model which incorporated both a relatively rapid decreasing
thermal SB component and a slower evolving non-thermal component.

Recently, more papers were published suggesting evolutionary schemes
which incorporate both SB and AGN.  \cite{2006ApJ...637..104K}
presented a sample of ULIRGs with type I Seyfert nuclei and showed
with both observations and modeling that this type of galaxies can be
powered by both a SB and a black hole (BH) with super-Eddington
accretion. They suggest a scheme in which SB-powered ULIRGs and QSOs
are two stages in the evolution of the host galaxy. The host will
start as a SB-driven ULIRG, with only a small BH. Over time, the SB
fades and the BH will grow into a super massive BH (SMBH), which will
dominate the energy output of the host. By that time the host has
become a QSO.  Similar arguments are made by
\cite{2006ApJS..163....1H} and \cite{2005astro.ph.11157H}.  On the
other hand, \cite{2005ApJ...635L.121K}, \cite{2005MNRAS.364.1337S} and
\cite{2005astro.ph.11157H} present schemes where the outflow of a
super-Eddington accreting SMBH drives into the surrounding ISM,
creating bubbles in which the gas cools and stars are formed. It is
clear that detailed modeling is necessary to determine the evolution
of ULIRGs and their engine(s).

\begin{figure}
  \centering
  \includegraphics[angle=0,width=\hsize]{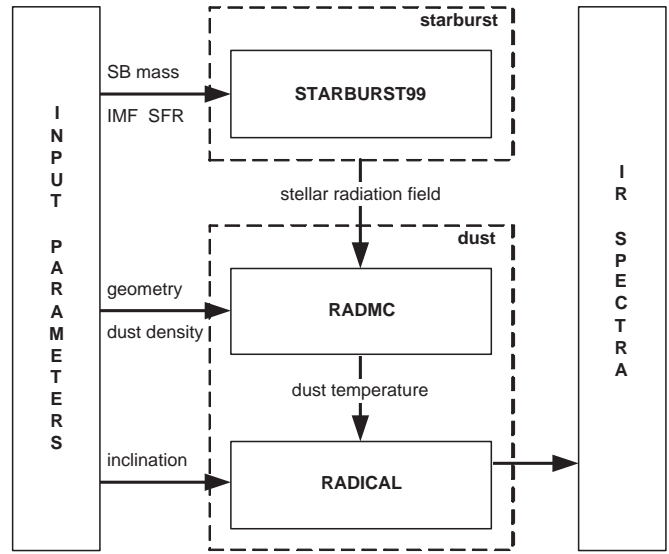}
  \caption{This flowchart shows the computational setup of the
  simulation. First the starburst properties are
  calculated using STARBURST99 (Sect. \ref{sec:starburst99}) and the
  results of these calculations are used as input for the dust
  calculations, which are performed by RADMC and RADICAL
  (Sects. \ref{sec:radmc} and \ref{sec:radical}).}
  \label{fig:code-flow}
\end{figure}

In this paper, we focus on massive star formation as a central
engine. In SB-powered ULIRGs, that are at sufficiently low
redshift for their internal structure to be resolved, the great
majority of the IR emission is found to originate from sub-kpc
dusty regions within merging systems of galaxies
\citep[e.g.][]{1998ApJ...507..615D}. These dusty SB galaxies are
an important class of objects. About 25\% of the high-mass star
formation within 10 Mpc distance from us occurs in just four SB
galaxies \cite[\object{M82}, \object{NGC253}, \object{M83},
\object{NGC4945};][]{1998ASPC..148..127H}.  Even though these
galaxies create vast amounts of stars, the time scale of this
formation is short.  Near-IR imaging spectroscopy in \object{M82},
\object{IC342}, and \object{NGC253} indicates that in the
evolution of these galaxies there are several episodes of star
formation activity, with timescales of around $10^7$ to $10^8$
years \citep{2000ARA&A..38..761G}.  The relatively low efficiency
of the energy production of stars ($E\sim 10^{-3} {\rm
M}_{\star}{\rm c}^2$) and the large energy output (up to $\sim
10^{61}$ ergs for an ultra-luminous starburst like
\object{Arp220}) yield a production $10^8$ to $10^{10}$ M$_{\sun}$
~of stars per burst. Combined with the short timescales, this
leads to star formation rates (SFRs) ranging from 10 up to 1000
M$_{\sun}$ per year \citep{1998ASPC..148..127H}. Numerical
simulations confirm this picture.  \cite{2006ApJ...637..255J} show
that during a typical merger event, there are several short
periods of star formation at a very high rate.

The goal of this paper is to present a model for these dust-enshrouded
star formation regions and to study and explain the behavior of the
broad band IR continuum properties of starburst ULIRGs
during their evolution.  The model consists of a number of existing
codes, which are combined into one.

In future work,  we will also investigate the spectral features in
the IR regime, which will provide more diagnostics to determine the
source of the activity in ULIRGs
\citep[e.g.][]{1999RvMP...71..173H,2000ARA&A..38..761G,2003PhDT........18S}.
We also intend to extend the model with a molecular environment, in
order to further constrain the physical parameters of the cores of
active galaxies \citep{2005A&A...436..397M}. We will also
investigate the similarities and differences between SB dominated and
AGN dominated ULIRGs and will investigate the possibility of an
evolutionary connection between the two.

The structure of this paper is as follows: in Sect. \ref{sec:model}
the model is discussed. The parameter space and the simulations are
presented in Sect.  \ref{sec:simulations} and the results of these
simulations in Sect.  \ref{sec:results}. In the last section the
results are discussed and suggestions for future work are made.

\section{The model}
\label{sec:model}
The ``physical'' setup of our model is shown in Fig. \ref{fig:setup}.
A star formation region is placed in the center of our model,
surrounded by a dusty environment, which is responsible for the
re-emission of the stellar UV photons in the IR.
The same scheme can be seen in Fig. \ref{fig:code-flow}, which shows
the computational setup of the model. The first step in the
simulations is the calculation of the spectral energy distribution
(SED) of the star formation region using STARBURST99. The surrounding
dust is modeled using RADMC and RADICAL, which use the stellar SED to
calculate the dust temperature and the emergent IR radiation. The used
codes will be briefly discussed and the interested reader is referred
to the listed references for further details.

\subsection{STARBURST99}
\label{sec:starburst99}
As mentioned above, the stellar properties of the starburst are
calculated using STARBURST99\footnote{The code and pre-calculated data
  sets are publicly available at
  http://www.stsci.edu/science/starburst99/} (hereafter
SB99), which is a set of stellar models, optimized to reproduce
properties of galaxies with active star formation and their evolution
through time.  In this paper only a brief summary of the model is
given. More information on the data produced by SB99 can be found in
\cite{1999ApJS..123....3L}. Detailed information on the computational
techniques can be found in \cite{1992ApJ...401..596L} and
\cite{1995ApJS...96....9L}.

The calculations can roughly be divided into three stages.  First
an ensemble of stars is formed according to the available gas mass
and the initial mass function (IMF), either in one instantaneous
burst (i.e., the duration of the burst is short compared to the
stellar life time) or continuous with a constant star formation
rate (SFR). Because these two cases represent the extremes of star
formation, realistic scenarios will lie somewhere in between
\citep{1995ApJS...96....9L,1992ApJ...401..596L}.

After creating the stellar population, evolutionary models are used to
calculate the time dependency of the stellar properties, such as
stellar mass, luminosity, spectra, effective temperature, and chemical
composition.

In the last stage of the calculation the integrated properties of the
star formation region are computed by determining the stellar number
densities, assigning all properties to these stars and integrating
over the entire population \citep{1995ApJS...96....9L}.

\subsection{RADMC}
\label{sec:radmc} The output SEDs of the SB99 calculations are
used to irradiate the dusty environment. In order to do this the
SB99 spectrum must be re-binned to the number of channels the dust
simulation codes can handle. The re-binning of the SB99 output is
performed in such a way that both the UV/optical part of the
spectra (which provides the energy for the dust heating) and the
IR regime (which is the part of the spectrum of interest) are well
covered.  Since the SB99 output SED ranges from 100\AA~(extreme
UV) up to wavelengths around $10^6$\AA~(100 $\mu$m; FIR) and the
number of channels is limited, the resolution of the re-binned SED
is low.  Therefore, the re-binning is performed in such a way that
the IR part of the SED has a somewhat higher resolution, so that
the 10$\mu$m, 20$\mu$m, 60$\mu$m and 100$\mu$m bands are included,
in order to compare the resulting IR data with IRAS
observations\footnote{In the SB99 output SED, 12 and 25 $\mu$m are
not available. Instead, 10 and 20 $\mu$m are used, which are good
approximations, given the low resolution of the data and the
simple dust model.}. Because, in this work, we are focusing on
defining a physical framework for an evolutionary model and not on
the details, and also taking the low spectral resolution and the
simple dust model into account, we did not compare our results
with more recent ISO or Spitzer data. Readers who are interested
in more detailed models which are not taking evolution into
account are refered to, for instance, \cite{1995MNRAS.273..649E},
\cite{2003A&A...404....1V}, \cite{2006MNRAS.366..767F} and
references there in.

The physical dust environment is simulated by creating a dust density
grid with a central cavity, which is the region surrounding the
central SB where the dust is destroyed and/or blown away by the star
formation activity. Since we also want to explore
the influence of the dust geometry on the emergent SEDs in our
simulations, a 3D grid is needed. This, however, would lead to long
calculation times. In order to limit this problem, the grid is chosen
to be axisymmetric. This way the grid is reduced to two dimensions
again, but still allows for various astrophysically interesting dust
distributions, like shells, tori and disks.

RADMC\footnote{For more information on RADMC, its availability and
related publications see
  http://www.mpia-hd.mpg.de/homes/dullemon/radtrans/radmc/
}
is a 2-D Monte-Carlo radiative transfer code for axisymmetric
circumstellar disks and envelopes. It is based on the method of
\cite{2001ApJ...554..615B}, but with several modifications
to produce smoother results with fewer photon packages.

\begin{table*}[ht!]
  \begin{minipage}[t]{\textwidth}
    \renewcommand{\footnoterule}{}  
    \caption{This table lists the parameters used for the
      simulations. The top half of the table shows the default
      simulation and the simulations done in the second stage. In this
      stage the large scale parameters were varied. The bottom half
      shows the simulation from the third stage, where the influence
      of the second dust component was investigated.  All simulations
      were calculated for inclinations 0, 30, 60 and 90 degrees. The
      names chosen for the simulations represent the parameter
      investigated in that simulation. Parameters not listed for a
      simulation were not used in that simulation.}
    \label{tab:sim-setup}
    \centering
    \begin{tabular}{llllllllllll}
      \hline
      \hline
      name & ${\rm M_{SB}}$           & $\alpha$        & $\tau$  & $\eta$ &SFR           & $r_0$&$r_1$&$dr_1$&$\eta_1$&$\tau_2$& dust\\
           & $\log{} {\rm M}_\odot$   & $M^{-\alpha}$   &         &  0-1 &  M$_{\sun}$/yr & pc & pc & pc&0-1&&\\
      \hline
      DEFAULT      &9  & 2.35 & 10  & 0.3 &&&&&&& grey torus  \\ 
      \noalign{\smallskip}
      LOWMASS      &8  & 2.35 & 10  & 0.3 &&&&&&& grey torus  \\ 
      HIGHMASS     &10 & 2.35 & 10  & 0.3 &&&&&&& grey torus  \\ 
      \noalign{\smallskip}
      LOWALPHA     &9  & 1.3  & 10  & 0.3 &&&&&&& grey torus   \\ 
      HIGHALPHA    &9  & 3.3  & 10  & 0.3 &&&&&&& grey torus   \\ 
      \noalign{\smallskip}
      LOWTAU       &9  & 2.35 & 1   & 0.3 &&&&&&& grey torus  \\ 
      HIGHTAU      &9  & 2.35 & 100 & 0.3 &&&&&&& grey torus  \\ 
      \noalign{\smallskip}
      COVERED1     &9  & 2.35 & 10  & 0.5 &&&&&&& grey torus  \\ 
      COVERED2     &9  & 2.35 & 10  & 0.7 &&&&&&& grey torus  \\ 
      \noalign{\smallskip}
      CONTINUOUS   &   & 2.35 & 10  & 0.3 & 100  &&&&&& grey torus  \\ 
      \noalign{\smallskip}
      \hline
      \noalign{\smallskip}
      30pcTORUS    &9  & 2.35 & 10  & 0.3 && 30 &&&&& grey torus \\
      50pcTORUS    &9  & 2.35 & 10  & 0.3 && 50 &&&&& grey torus \\
      \noalign{\smallskip}
      10pcSHELL    &9  & 2.35 & 10  & 0.3 &&& 10 & 10 & 1 & $\tau + \tau_2$    = 10 \footnote{The total column of both the stellar and the torus dust is scaled to an optical depth of 10. See Sect.~\ref{sec:extra-dust-component} for more information.}& grey torus and grey stellar   \\
      10*10pcSHELL &9  & 2.35 & 10  & 0.3 &&& 10 & 10 & 1 & $\tau + 10*\tau_2$ = 10  \footnote{First the density of the stellar dust is multiplied by a factor of 10, then the total column of both the stellar and the torus dust is scaled to an optical depth of 10. See Sect.~\ref{sec:extra-dust-component} for more information.}& grey torus and grey stellar  \\
      \noalign{\smallskip}
      FINAL        &9  & 2.35 & 10  & 0.3 &&& 10 & 10 & 1 &0.1& grey torus and non-grey stellar  \\

      \hline
  \end{tabular}
 \end{minipage}
\end{table*}

The method introduced by \cite{2001ApJ...554..615B} is to divide the
luminosity of the radiation source (in this case the re-binned output
spectrum of SB99) into equal-energy, monochromatic ``photon packets''
that are emitted stochastically by the source. The packets are then
traced to random interaction locations on the dust grid, determined by
the optical properties and density distribution of the dust.  In every
iteration step, the number of packets absorbed in each of the grid
cells is recorded and the total amount of injected energy is used to
calculate the new temperature of that cell. In order to ensure energy
conservation and radiative equilibrium, the absorbed energy must be
re-emitted immediately. The frequencies of the new packets are
determined by the new grid cell temperature. Local thermal equilibrium
(LTE) is assumed, so the amount of emitted energy equals the absorbed
energy.

Because all the injected packets eventually escape and so does all
the injected energy, the total energy is automatically conserved.
Another advantage of this approach is that the simulation does not
have some sort of convergence criterion to meet, so that the only
source of error in the calculation is the statistical error
inherent to Monte Carlo simulations.

One problem with the \citeauthor{2001ApJ...554..615B} method is
that it produces very noisy temperature profiles in regions of low
optical depth, and requires a large number of photons ($N \sim
10^7$) for a smooth SED. These disadvantages have been solved in
RADMC by treating absorption partly as a continuous process (see
\citealt {1999A&A...344..282L} for more information), and using
the resulting smooth temperature profiles with a separate
ray-tracing code (RADICAL) to produce images and SEDs. These
images and SEDs have a low noise level even for relatively few
photon packages ($N \sim 10^5$).  This improved
\citeauthor{2001ApJ...554..615B} method works well at all optical
depths, but may become slow in cases where the optical depth is
very large \citep[$\tau_\nu \sim$ 1000;
see][]{2004A&A...417..793P}; this situation is not reached in our
simulations.

\subsection{RADICAL}
\label{sec:radical}

RADICAL\footnote{For more information on RADICAL, its availability and
related publications see
http://www.mpia-hd.mpg.de/homes/dullemon/radtrans/radical/} produces
the final spectra, by solving the radiative transfer equation on the
entire grid, using the dust temperature and the SED of the central
source to calculate the source function.

First an image is created by integrating the source function along
rays through the medium. Each ray represents one pixel of the
image. Spectra can be derived by making images at a range of
frequencies, and integrating these over an artificial detector
aperture.

Resolution problems may occur, since the source spans a large radial
range. The central parts of the image are often much brighter than the
rest, but cover a much smaller fraction of the image.  The spectrum
may therefore contain significant contributions of flux from both the
central parts and the outer regions of the image. Unless the image
resolves all spatial scales of the object, the spectra produced in
such a way are unreliable. Therefore, rather than arranging the pixels
over a rectangle, as in usual images, they are arranged in concentric
rings. The radii of these rings are related to the radial grid points
of the transfer calculation. To resolve the central source, some extra
rings are added.  Using these kind of images, all relevant scales are
resolved, while using only a fairly limited number of pixels
\citep{2000A&A...360.1187D}.

\begin{figure*}
  \resizebox{0.3\hsize}{!}{\includegraphics[bb=54 360 450 720]{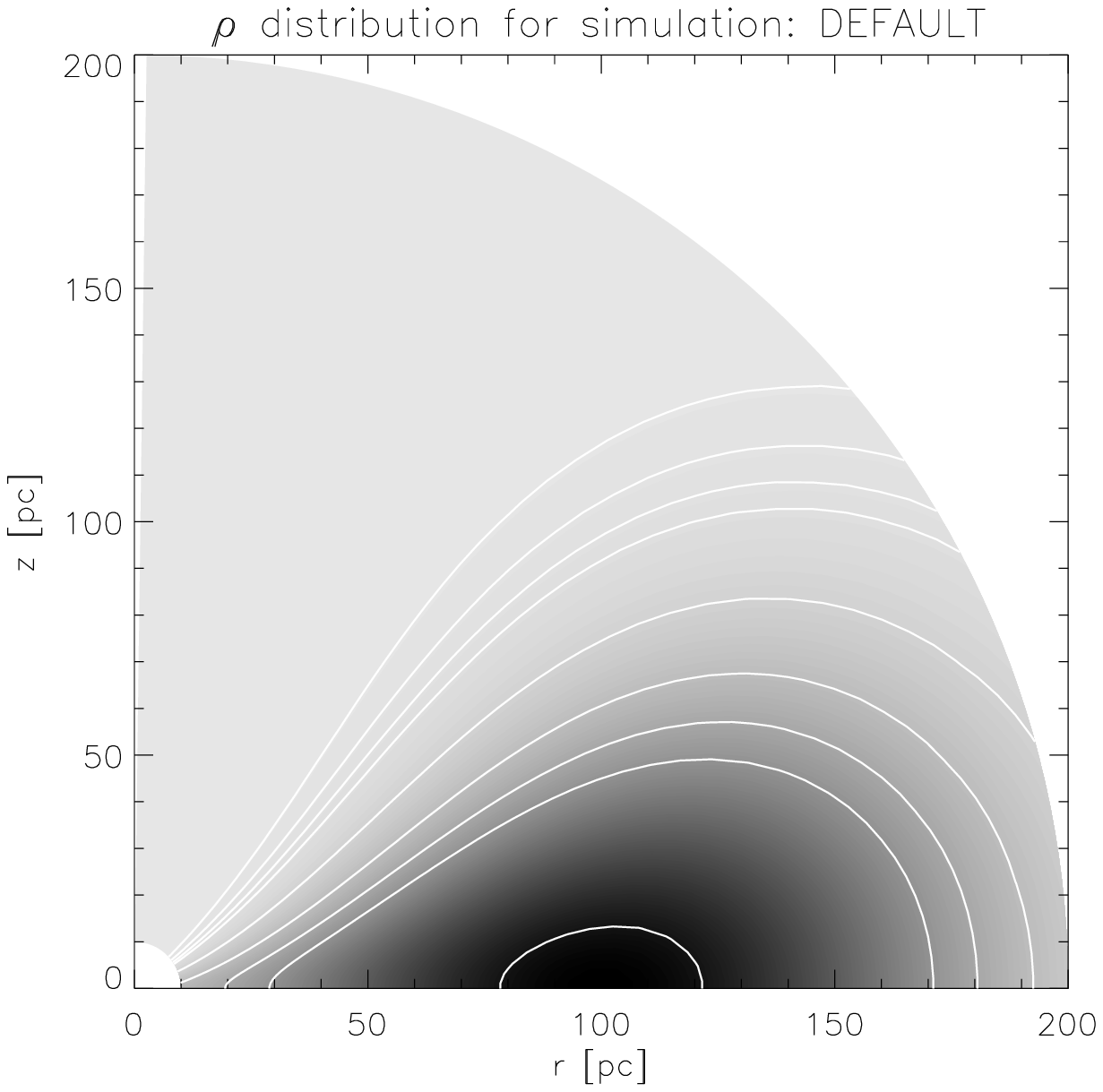}}
  \hspace{0.03\hsize}
  \resizebox{0.3\hsize}{!}{\includegraphics[bb=54 360 450 720]{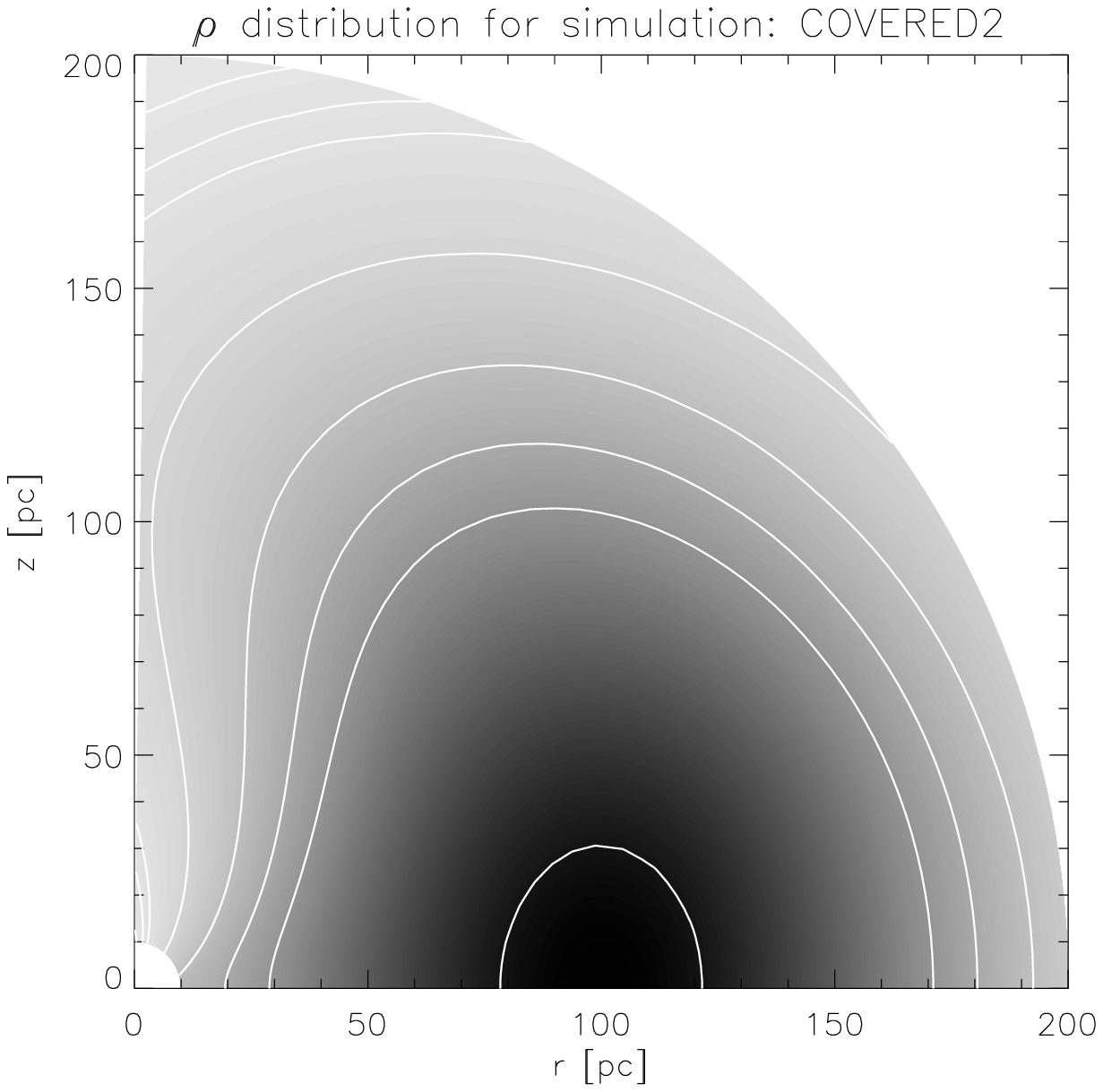}}
  \hspace{0.03\hsize}
  \resizebox{0.3\hsize}{!}{\includegraphics[bb=54 360 450 720]{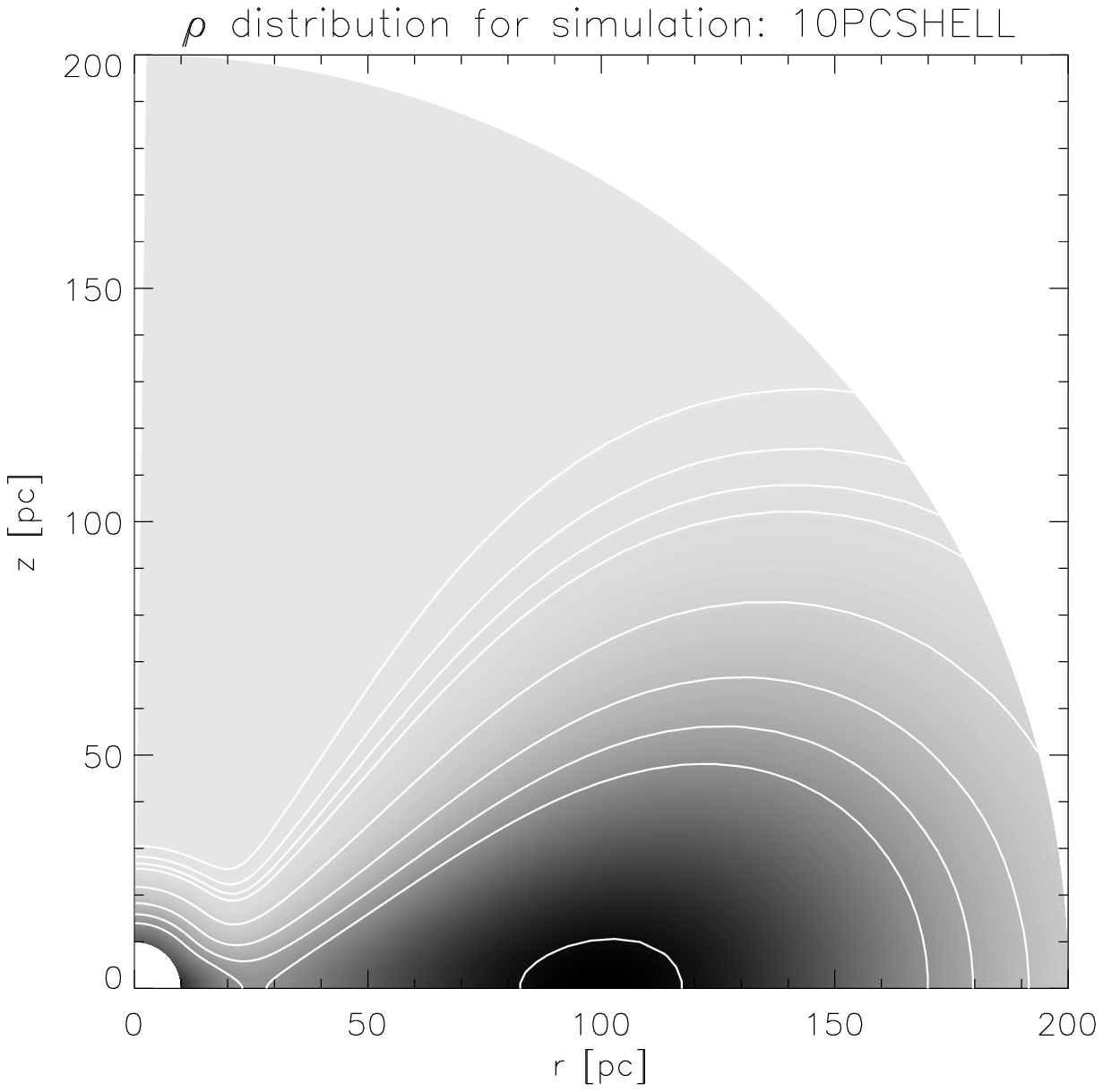}}

  \vspace{0.03\hsize}

  \resizebox{0.3\hsize}{!}{\includegraphics[bb=54 360 450 720]{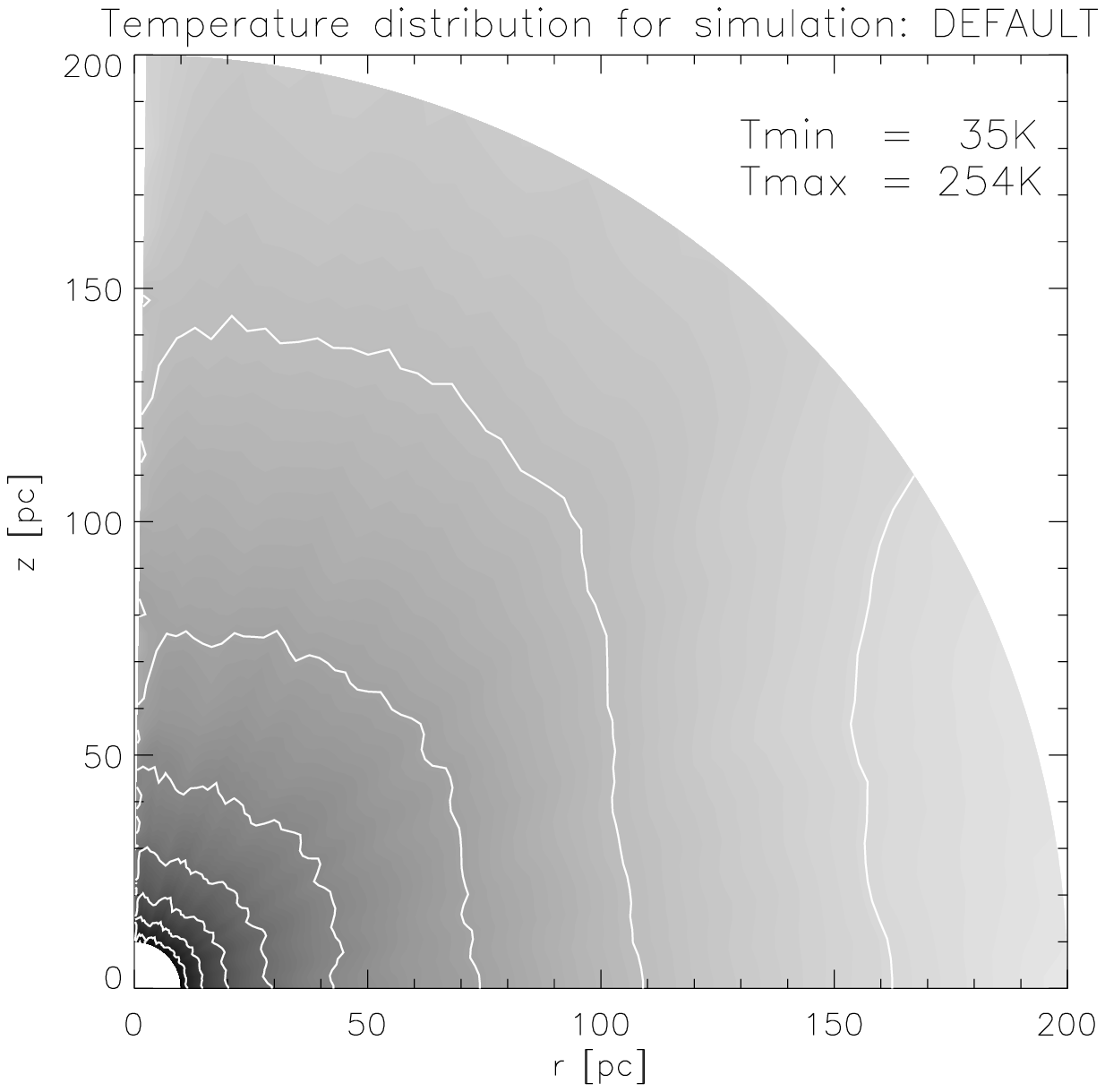}}
  \hspace{0.03\hsize}
  \resizebox{0.3\hsize}{!}{\includegraphics[bb=54 360 450 720]{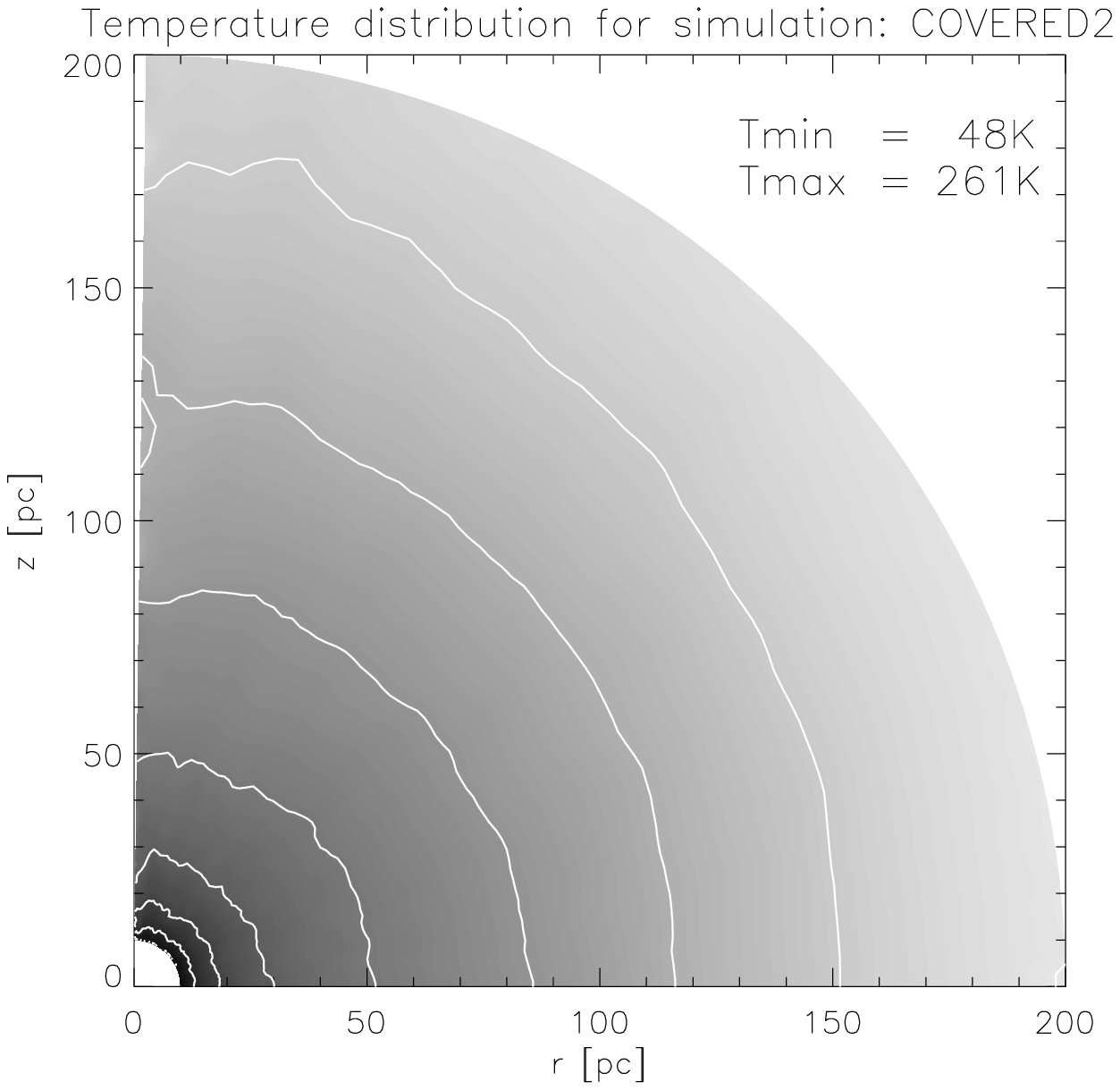}}
  \hspace{0.03\hsize}
  \resizebox{0.3\hsize}{!}{\includegraphics[bb=54 360 450 720]{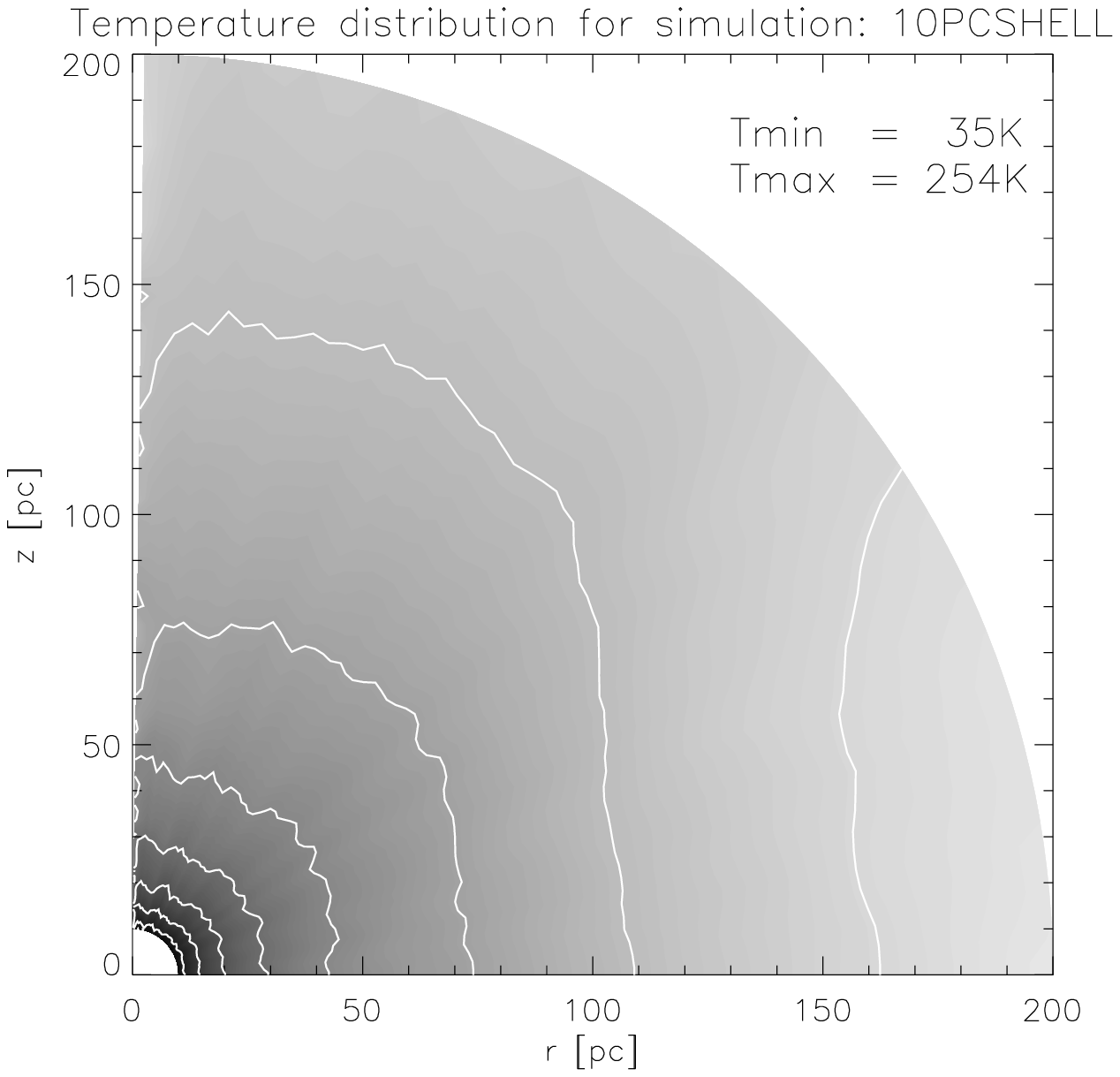}}

  \caption{These figures show three examples of the dust and
  temperature distributions resulting from the simulations. The top
  panels show the dust density in grey scale, where dark colors
  indicate regions with high densities and light colors are low
  density regions. Also some contours are added to enhance the shape
  of the density profile.  The bottom panels show the temperature
  distribution. In these panels, dark colors indicate high
  temperatures and light colors cooler regions. The minimum and
  maximum temperature is printed in the upper right corner of the
  panels. Contours are drawn from 50 to 250K in steps of 25K.  The
  data in these plots was smoothed over 3 pixels to make the general
  temperature distribution more clear. The left panels show the
  default simulation, the middle ones a setup with $\eta = 0.7$ and
  the right ones a default distribution with an extra 10 pc shell of
  grey dust.  }
  \label{fig:dust-T}
\end{figure*}

\section{The simulations}
\label{sec:simulations}

In this section, the simulation setup and the parameter space are
discussed.  As shown in Fig. \ref{fig:setup}, there are 6 parameters,
which control the simulation: the combined mass of the stars in the
starburst ${\rm M_{SB}}$, the star formation rate (SFR), the IMF slope
$\alpha$, the closing angle of the dust distribution $\eta$, the dust
column expressed in the edge-on V-band optical depth $\tau$ and the
observation declination $\delta$.

This parameter space has been explored, starting with values chosen
either from literature or at random. In order to verify the results of
the simulations, a reference data set was made using IRAS data of
starburst galaxies taken from the extended 12 micron galaxy sample
\citep{1993ApJS...89....1R} and the IRAS Bright Galaxy Sample
\citep{1991ApJ...378...65C}.

From the simulations a default set is chosen, which fits the data
best. This set will be referred to as DEFAULT. This set was chosen as
the starting point for a more systematic sweep of the parameter
space. Each parameter was both increased and decreased with respect to
the default value. Exceptions are the closing angle $\eta$, the
inclination $\delta$ and the SFR. Both permutations of $\eta$ are
higher than the default value. The inclination was varied in all
simulations: each setup was calculated for $\delta$ = 0, 30, 60 and 90
degrees. The SFR was only used in few simulation. In the other
simulations, all stars are formed instantaneous.  The entire parameter
space is listed in Table \ref{tab:sim-setup}.

\subsection{STARBURST parameters}
\label{sec:starburst-parameters}%
The SB in the center of the simulation has a large number of
parameters. Only three are chosen as variables, in order to reduce
the amount of data and to avoid unnecessary redundancy. These
parameters are also expected to have the largest influence on the
final results. The fixed parameters are listed in Table
\ref{tab:SB99-parameters}.

The variable parameters are the total mass of the stars involved in
the starburst (${\rm M_{SB}}$), the star formation rate (SFR) and the
IMF slope\footnote{The IMF is defined as: $\Psi(M) \propto
M^{-\alpha}$.} ($\alpha$).  These were chosen because the large
luminosities of ULIRGs can be explained by either a high total burst
mass, a high SFR or a lower IMF slope exponent. Adding more stars or
increasing the SFR will scale the entire spectrum, whereas decreasing
the IMF slope increases the relative number of heavy stars, leading to
more UV radiation and therefore more heating flux \citep[as suggested
in][]{1998ASPC..148..127H}.

\begin{table}
  \begin{minipage}[t]{\columnwidth}
    \caption{This table lists the parameters which determine the
      simulations, but were not varied. The upper part of the table lists the SB99
      parameters, the lower part the RADMC parameters.}
    \label{tab:SB99-parameters}
    \centering
    \renewcommand{\footnoterule}{}  
    \begin{tabular}{lll}
      \hline
      \hline
      name & value & \\
      \hline
      IMF upper mass limit     & 100  & M$_{\sun}$ \\
      IMF lower mass limit     & 1    & M$_{\sun}$ \\
      supernova cut-off mass   & 8    & M$_{\sun}$ \\
      black  hole cut-off mass & 120  & M$_{\sun}$ \\
      \noalign{\smallskip}
      initial time             & 0.01 & Myr\footnote{
    For numerical reasons SB99 cannot start the simulation at t=0.}\\
      time step                & 0.1  & Myr\\
      end time                 & 20   & Myr\\
      \noalign{\smallskip}
      wind model                      & \multicolumn{2}{l}{theoretical\footnote{
      See \citet{1992ApJ...401..596L} and \citet{1999ApJS..123....3L} for more information.}}\\
      atmosphere model                & \multicolumn{2}{l}{Pauldrach/Hillier$^b$}\\
      metallicity of UV line spectrum & \multicolumn{2}{l}{solar}\\
      \noalign{\smallskip}
      \cline{1-3}
      \noalign{\smallskip}
      $r_0$                     & 100  & pc\footnote{In the third stage, some simulations were made with a different $r_0$ }\\
      $dr_0$                    & 50   & pc\\
      \noalign{\smallskip}

      \hline
    \end{tabular}
  \end{minipage}
\end{table}

\subsection{Dust setup}
\label{sec:dust-setup}
The dust distribution is made by creating a polar grid, with a density
on each grid point. The shape is created by introducing a Gaussian
distribution in both the $r$- and $\theta$-direction:
\begin{equation}
\rho\left(r,\theta\right) =
\exp{\left[-\frac{1}{2}\left(\frac{\pi/2-\theta}{\eta}\right)^2
-\frac{1}{2}\left(\frac{r-r_0}{dr_0}\right)^2\right]}~,
\label{eq:41}
\end{equation}
where $\rho\left(r,\theta\right)$ is the dust density on each grid
point, $r$ is the radial direction, $\theta$ the angular direction,
$r_0$ the radius of the dust geometry, $dr_0$ the radial scale size
and $\eta$ the closing angle. This density distribution is then fixed
by scaling the edge-on optical depth (obtained by integration along
the line of sight) to the desired V-band optical depth $\tau$.  Three
examples of resulting density distributions are plotted in
Fig. \ref{fig:dust-T}.  From this two-dimensional grid, a
three-dimensional grid is made by first mirroring the grid on the
$r$-axis and then rotating this semi-circle around the z-axis.

Two dust components are created: a large scale component distributed
in a thick disk/torus (here after called torus dust) and a small scale
shell (called stellar dust). The torus dust distribution is created by
using an $r_0$ of 100 pc and  closing angles ($\eta$) between 0.3 and
0.7. The stellar dust distribution is made by taking an $r_0$ of 10 pc
and an $\eta$ of 1, making it a closed shell.

Besides the density and geometry, also the optical parameters of the
dust must be specified.  The dust used in the simulations has a grain
size distribution $n(a) \propto a^{-3.5}$
\citep{1977ApJ...217..425M}. The torus dust is assumed to be optically
grey, i.e., radiation of all frequencies is absorbed in the same way
and re-radiated using a black body profile. The stellar dust is much
closer to the SB and will be hotter. In order to model the expected
richer spectral properties of this component, the stellar dust is
assumed to be composed of graphite, silicates and amorphous carbon
\citep[e.g.][]{1984ApJ...285...89D}.\\

\section{Results}
\label{sec:results}

The code produces four  broadband IR fluxes as a function of
time. These fluxes are used to calculate the IR luminosity (${\rm
L_{IR}}$) following the definitions given in
\citet{1996ARA&A..34..749S} and \citet{1998ApJS..119...41K}.

The simulations were carried out in three stages. First a default
simulation was made. In the second stage the large scale parameters,
like the starburst and the torus dust, are explored, which influence
the long wavelength radiation. The last stage focuses on the stellar
dust, which affects the short wavelength emission.

\subsection{Stage 1 - The default simulation}
\label{sec:defaults}

\begin{figure*}[!ht]
  \centering
  \resizebox{.8\textwidth}{!}{\includegraphics[angle=0]{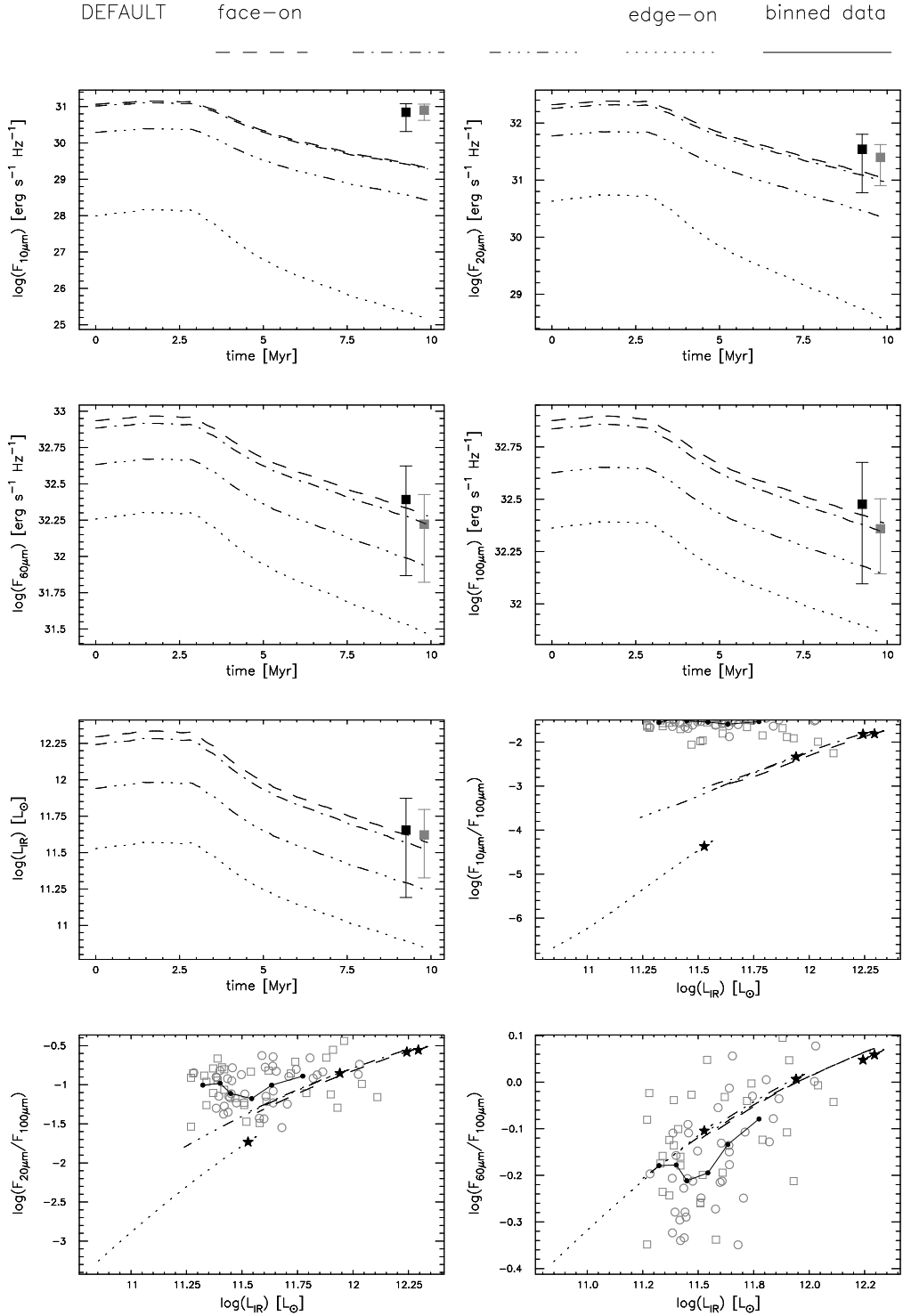}}
     \caption{These panels show the infrared results of the simulation
              DEFAULT. The top 4 panels show the 10, 20, 60 and
              100$\mu$m flux as a function of time. The fifth panel
              shows the calculated infrared luminosity (${\rm
              L_{IR}}$) through time. The last three panels give a
              more observational interpretation of the data. In these
              panels three IR colors (10$\mu$m/100$\mu$m,
              20$\mu$m/100$\mu$m and 60$\mu$m/100$\mu$m) are plotted
              versus the IR luminosity. To indicate the evolution
              direction, the start of the track is marked with a
              star. All axes are logarithmic except the time axes. The
              observational inclinations are indicated by the line
              style: from face-on in solid, through dashed and
              dash-dotted to edge-on indicated with dotted lines.  The
              two reference data sets are added with markers. In the
              first 5 panels, the error markers represent the mean of
              these data sets and a 1 $\sigma$ deviation. The grey
              marker is data from the 12 micron galaxy sample
              \citep{1993ApJS...89....1R} and the black one from the
              IRAS Bright Galaxy Sample
              \citep{1991ApJ...378...65C}. In the last three panels,
              the data from the 12 micron galaxy sample is indicated
              by circles and the data from the IRAS Bright Galaxy
              Sample by squares. The black dots connected by a solid
              line show the reference data binned per 10 points.}
      \label{fig:defaultAll}
\end{figure*}

The results of the default simulation from the first stage are plotted
in Fig. \ref{fig:defaultAll}.  The first 4 panels show the results for
the four IRAS bands.  As can be seen the simulation results match the
reference set quite well. Only the shortest wavelength at 10$\mu$m
seems to be a poor fit. This is due to the fact that in this stage we
have not included the stellar dust yet (see
Sect. \ref{sec:other-simulations} for the stellar dust). This dust
component will radiate mostly at short wavelengths.

The calculated ${\rm L_{IR}}$, which is plotted in the fifth panel,
also covers the range observed in galaxies. At the peak of its
activity the luminosity of the starburst exceeds $10^{12}$ L$_{\sun}$,
making it a ULIRG. It remains that luminous for about 3 Myr, after
which it slowly dims and eventually reaches the LIRG regime.

In the last three panels of Fig. \ref{fig:defaultAll}, three IR
colors (10$\mu$m/100$\mu$m, 20$\mu$m/100$\mu$m and
60$\mu$m/100$\mu$m) are plotted versus the IR luminosity. These
color-luminosity diagrams (CLDs) show results similar to the IRAS
fluxes. The 20$\mu$m/100$\mu$m and 60$\mu$m/100$\mu$m results
match the observations, whereas the 10$\mu$m/100$\mu$m is a poor
fit, due to the absence of the stellar dust.

\begin{figure}
  \centering
  \includegraphics[angle=0,width=\hsize]{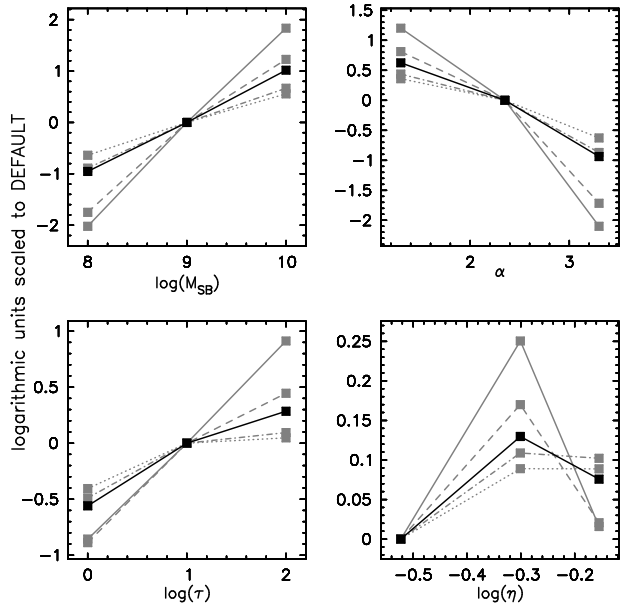}
  \caption{These panels give an overview of the simulation
  results. The four IRAS bands (in grey: 10$\mu$m solid, 20$\mu$m
  dashed, 60$\mu$m dash-dotted and 100$\mu$m dotted) and the IR
  luminosity (black solid) are shown as a function of the various
  parameters (top-left: ${\rm M_{SB}}$, top-right: $\alpha$,
  bottom-left: $\tau$, bottom-right: $\eta$).  The points represent
  the values at 2Myr, which is the time where the luminosity of the
  system is at its maximum. All data were scaled to the results of
  DEFAULT, in order to make an easy comparison. Note that the results
  for the SFR are not included in these plots. Also the observational
  inclination $\delta$ was not included: the data are taken from the
  face-on results.}
\label{fig:allResults}
\end{figure}
\begin{figure*}[ht!]
  \centering
  \includegraphics[angle=0,width=\hsize]{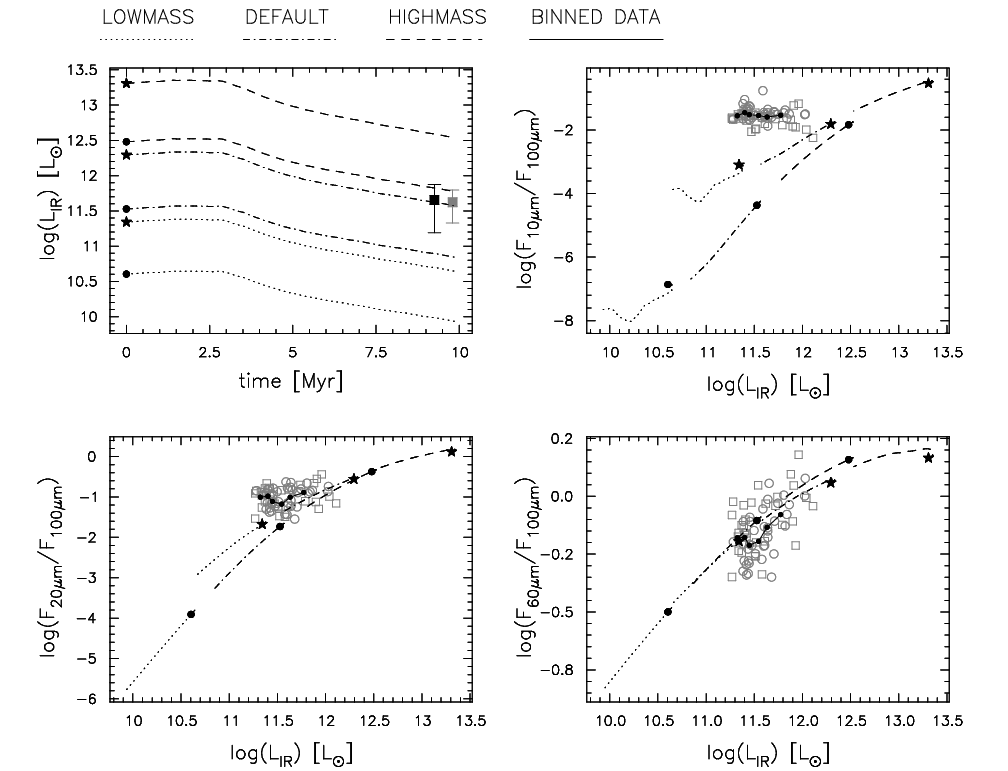}
  \caption{These panels show the influence of the starburst mass
  (${\rm M_{SB}}$) in the infrared results. The two deviations are
  compared to the default value. The simulation results for the lower
  deviation are indicated by dotted lines, the higher deviation by
  dashed lines and the default set is drawn in dash-dotted lines. The
  face-on observations are marked with a star, the edge-on results
  with a circle. These marks also indicate the start of the
  evolutionary track, to indicate the evolution direction. The panels
  are the same as the bottom four panels of Fig. \ref{fig:defaultAll}:
  the first panel shows the IR luminosity as function of time and the
  other three panels show three IR colors (10$\mu$m/100$\mu$m,
  20$\mu$m/100$\mu$m and 60$\mu$m/100$\mu$m) versus the IR
  luminosity. As in Fig. \ref{fig:defaultAll} the two reference data
  sets are added with markers. In the first panel, the error
  markers represent the mean of these data sets and a 1 $\sigma$
  deviation. The grey marker indicates data from the 12 micron galaxy sample
  \citep{1993ApJS...89....1R} and the black one from the IRAS Bright
  Galaxy Sample \citep{1991ApJ...378...65C}. In the last three panels,
  the data from the 12 micron galaxy sample is indicated by circles
  and the data from the IRAS Bright Galaxy Sample by squares. The
  black dots connected by a solid line show the reference data binned
  per 10 points.}
\label{fig:mass-results}
\end{figure*}

The default results exhibit a strong dependence on
inclination. The difference between edge-on and face-on ranges
from a factor of 3.2 at 100$\mu$m up to 1000 at 10$\mu$m, with an
average factor of 5.6 in the ${\rm L_{IR}}$. This large separation
at short wavelengths is caused by the dust distribution. As can be
seen in Fig. \ref{fig:dust-T}, the resulting temperature
distribution is not spherical. Along the edge-on line of sight the
optical depth is 10, whereas along the face-on line of sight the
star formation region is almost ``naked''.  This causes the areas
with lower temperature to become more flattened, compared to the
inner, high temperature areas. When this temperature distribution
is viewed edge-on, the cooler isothermal surfaces are seen at
their maximum surface area and when observed face-on, their
surface areas decrease rapidly. On the other hand, the inner high
temperature regions are roughly spherical and will therefore look
the same from all directions. The overall face-on spectrum will
thus have a larger contribution at short IR wavelengths, as
compared to the edge-on spectrum. This is also illustrated in the
lower right panel of figure \ref{fig:dust-T}, which shows the
temperature distribution of simulation COVERED2, in which the
cooler dust is more spherical and the inclination dependence of
the 10$\mu$m flux is an order of magnitude smaller. Another effect
of the dust distribution is that short wavelength radiation
emitted in the center will encounter much more dust when it is
traveling along the edge-on line of sight compared to face-on,
making the chance of it getting scattered or re-processed much
larger.

The effect of this inclination dependence can also be seen in
the three CLDs in figure \ref{fig:defaultAll}. At short wavelengths
(10$\mu$m/100$\mu$m and 20$\mu$m/100$\mu$m) the face-on observations
trace the data sets, whereas the edge-on results cover an entirely
different range. This effect will be discussed in more detail, when
the results of the other simulations are presented.

\begin{figure*}
  \centering
  \includegraphics[angle=0,width=\hsize]{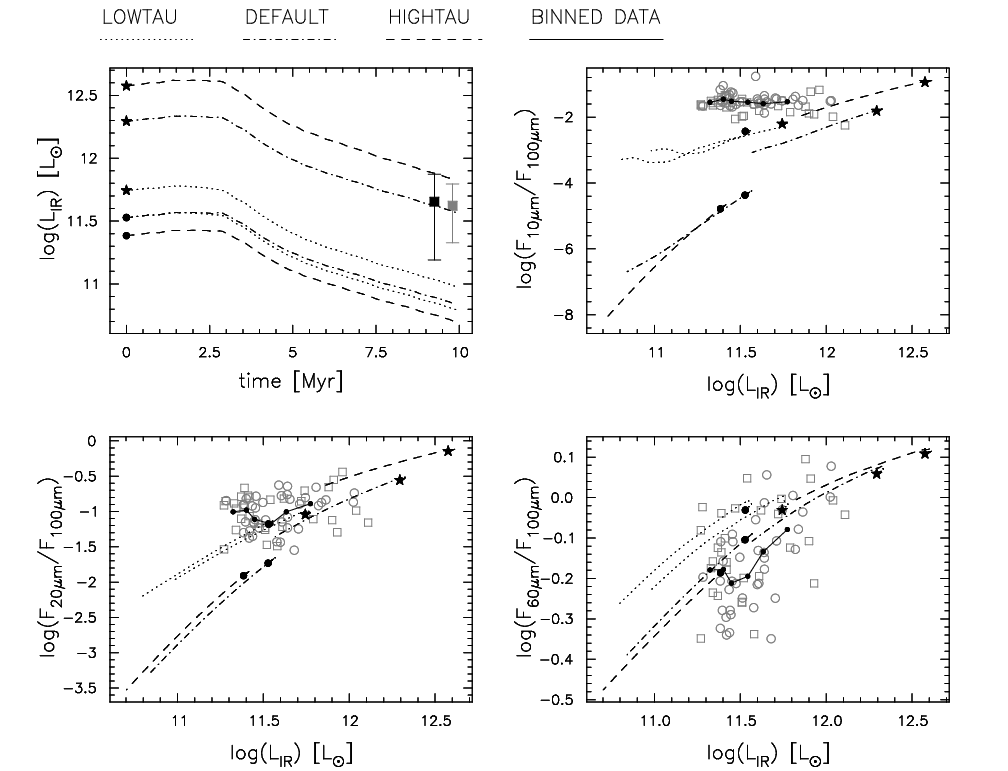}
  \caption{These panels show the influence of the dust optical depth
  ($\tau$) on the infrared results. The color coding is the same as in
  Fig. \ref{fig:mass-results}.}  \label{fig:tau-results}
\end{figure*}

\subsection{Stage 2 - Large scale parameters}
\label{sec:sweep-param-space}

In this section the parameters influencing the star formation and
the torus dust are explored.  Each parameter was varied twice, in
most cases ``in both directions'' from the default values, in
order to uncover systematic effects. The observational inclination
will not be discussed, since the effects of the inclination have
already been discussed above.  The figures which show the most
important effects have been included in this article. An overview
of the results is shown in Fig. \ref{fig:allResults}.  The rest of
the figures displaying all the results, including the four IRAS
fluxes, of the individual simulations and comparisons between them
are collected in Appendices A and
B respectively.

\subsubsection{Starburst mass}
\label{sec:starburst-mass}

The effect of the starburst mass on the SB99 spectra is quite
simple. Since this code works with a distribution of stars, integrated
to a certain mass, the spectra will scale proportional to the total
stellar mass. The influence on the dust is less straightforward,
however. When the energy input provided by the starburst increases,
the overall dust temperature will be higher and the flux at short
wavelengths will increase faster relative to the long
wavelengths. Also the IR luminosity is expected to increase with the
stellar mass, since the energy input increases.

The simulations confirm these expectations. Figure
\ref{fig:allResults}, which gives an overview of the results of the
simulations, clearly shows that the short wavelengths (grey solid and
dashed) increase much faster than the long wavelengths (grey
dash-dotted and dotted) and that ${\rm L_{IR}}$ (black) increases as
the mass increases.

The color-luminosity plots in Fig. \ref{fig:mass-results} seem to
reveal less trivial information. Assuming all the other parameters are
correct, a minimal SB mass of $10^9$~M$_{\sun}$ ~is needed in order to
create an IR luminosity higher than $10^{12}$ L$_{\sun}$. Also all
three simulations seem to lie on the same ``track'' in the CLDs, and
starburst mass and age have the same effect on the results. In other
words, a young, but relatively light starburst will, in IR colors,
have the same appearance as a massive, older starburst.

\subsubsection{IMF slope}
\label{sec:imf-slope}

When the IMF slope decreases, the number of massive stars in the
starburst will increase. These high mass stars are the main source of
UV radiation, and so the dust temperature will increase with a
shallower IMF slope. This means that varying the IMF slope would yield
results similar to increasing the starburst mass.

\begin{figure*}
  \centering
  \includegraphics[angle=0,width=\hsize]{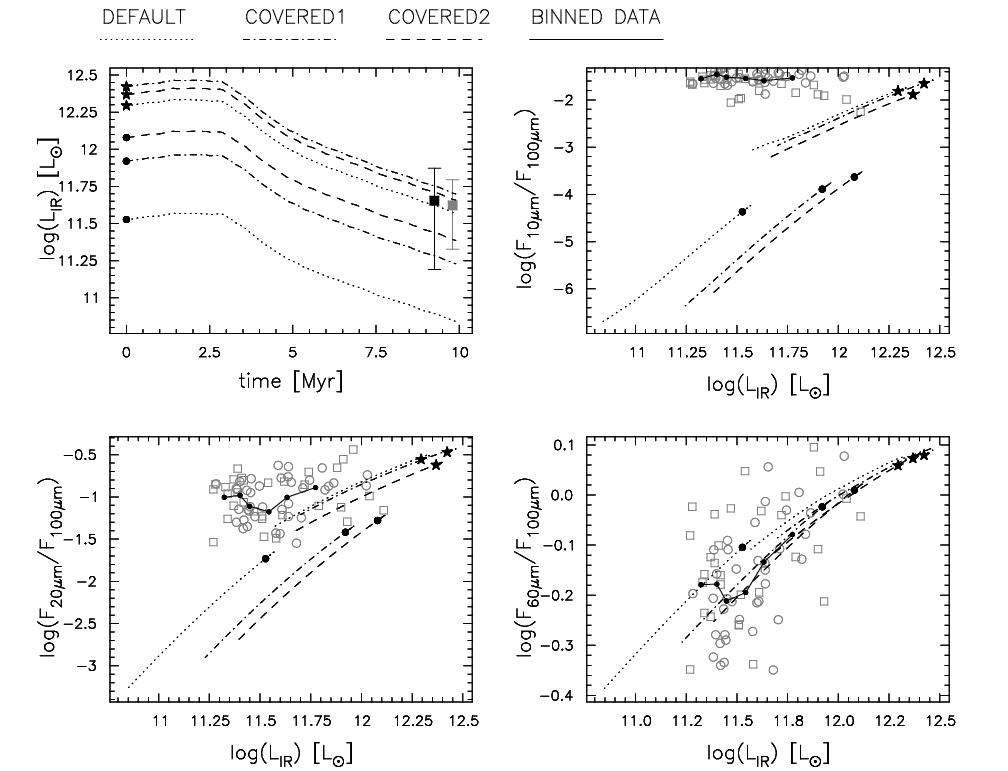}
  \caption{These panels show the influence of the closing angle
  ($\eta$) on the infrared results. The color coding is the same as in
  Fig. \ref{fig:mass-results}. Note: for this simulation both
  deviations are higher than the default value.}
  \label{fig:cover-results}
\end{figure*}

Figure \ref{fig:allResults} shows this is indeed the case. Note that
at first sight the effect seems to be reversed, but the IMF is defined
as $\Psi(M) \propto M^{-\alpha}$, and the axis is
reversed. Interestingly, the effect is somewhat more complicated than
for ${\rm M_{SB}}$. All fluxes scale proportionally to ${\rm M_{SB}}$,
but the effect is stronger when going from high $\alpha$ to
the default value, then when lowering it even further. This is caused
by the fact that the number of OB stars does not increase linearly
with $\alpha$.

Figure B.2 shows that although LOWALPHA starts
with higher values for the IR luminosity and the colors, after about
10 Myr the low IMF case cannot be distinguished from the default
simulation any more. This is due to the short life time of the high
mass stars.  Therefore, the same conclusion can be drawn for the IMF
slope as for the starburst mass: the data either represent young
starbursts with a normal IMF, or an older population of stars with a
shallow IMF slope.

\subsubsection{Dust column density}
\label{sec:dust-density}

Several effects have been identified when changing the total
amount of dust for a given energy input.

First, when the dust column is lower, the dust will be heated more
evenly, since the radiation can travel through the dust easier.
This will result in a smaller temperature gradient, with a higher
minimum temperature. The spectrum will become flatter, and have
less contribution from 10$\mu$m and 20$\mu$m. Conversely, large
dust columns lead to larger dust temperature gradients and steeper
spectra. This effect is shown in Figure \ref{fig:allResults},
where the variation of $\tau$ has effects similar to ${\rm
M_{SB}}$ and $\alpha$.

Second, due to the larger temperature gradient, the inclination
dependence  increases for larger $\tau$. In Figure
\ref{fig:tau-results}, this effect can be seen in the first panel,
where the edge-on and face-on results are almost the same for
LOWTAU, whereas the results for HIGHTAU differ by up to an order
of magnitude.

A third effect is the increase in the total IR luminosity ${\rm L_{IR}}$,
simply because there is more radiating dust and therefore more UV
photons can be converted into IR.

\begin{figure*}
  \centering
  \includegraphics[angle=0,width=\hsize]{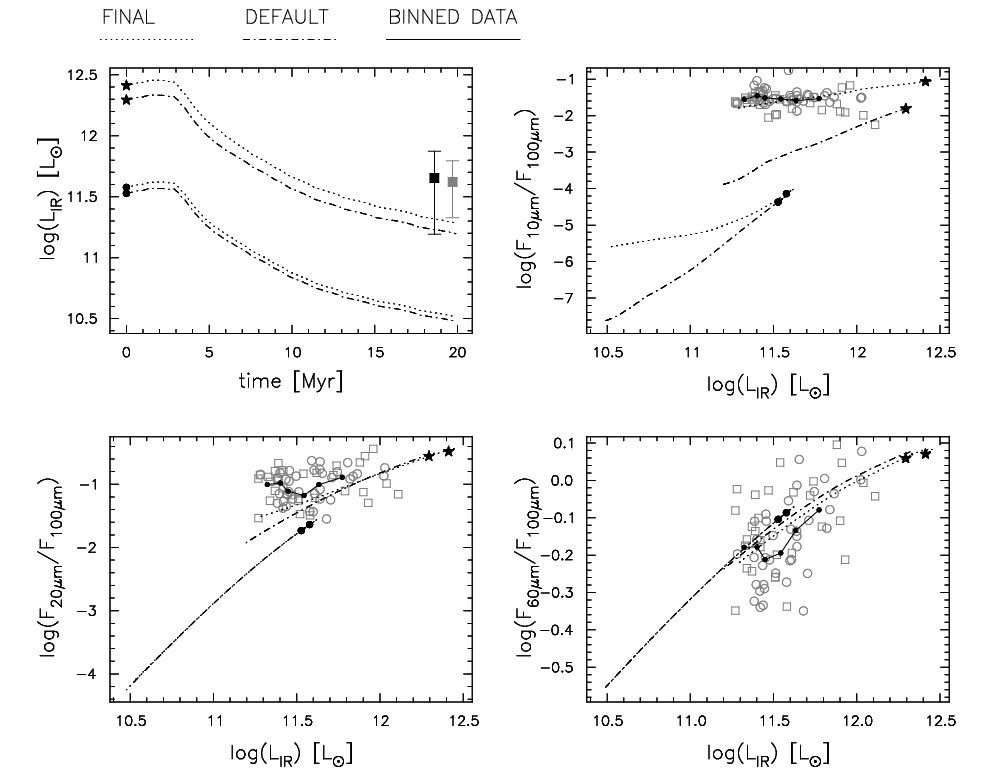}
  \caption{These panels show the influence of adding the stellar dust
  component on the infrared results. The color coding is the same as
  in Fig. \ref{fig:mass-results}.  Note: to show that the
  whole reference set is traced in the CLDs, 20 Myr is shown instead
  of 10.}
  \label{fig:final-results}
\end{figure*}

All these effects can be found in the color-luminosity diagrams.
When going from LOWTAU to DEFAULT especially ${\rm L_{IR}}$
increases, resulting in a shift to the right. But when $\tau$
increases even more, the color changes rapidly, and the track
moves to the upper right corner of the plot. Again, due to the
inclination effects, the edge-on results do not match the
reference sets. The edge-on results of LOWTAU are better, because
they suffer less from inclination effects, but they give worse
fits than the face-on DEFAULT results.

Like in the previous simulations, there is a degeneracy between $\tau$
and time. Young starbursts surrounded by moderate column density dust
have the same IR properties as older, more enshrouded starbursts.

\subsubsection{Closing angle}
\label{sec:closing-angle}

The effect of the closing angle is not as profound as that of the
other parameters. The only major effect is that the inclination
dependence effects are reduced, which is caused by a smaller
density gradient in the $\theta$ direction (also see Fig.
\ref{fig:dust-T}). The extreme case would be a simulation with
$\eta=1$, which is a shell of dust with no inclination dependence
at all.

Although the inclination effects are less in simulations with a higher
closing angle, the edge-on tracks in the color-luminosity tracks
of Fig. \ref{fig:cover-results} still do not fit the reference data.

The shape of the spectrum also changes when the closing angle changes,
but this is due to the fact that the total dust column changes,
since the edge-on optical depth is kept constant.

\subsubsection{Star formation rate}
\label{sec:star-formation-rate}

So far, all simulations were made under the assumption that the stars
were formed instantaneously, or, in other words, on a timescale much
shorter than the evolutionary timescale of the stars ($\lesssim$ 1
Myr). This is based on the assumption that there is a rapid build-up
of gas in the starburst environment, which at some point becomes
critically dense and ``ignites''. It also assumes that the stellar
winds and supernovae do not provide strong feedback, so that all gas
can be converted into stars on a short time scale.  However, the
assumption that the starburst environment is less violent and more
like normal star formation regions, with a steady inflow of gas, is
also valid and must be explored as well.

In order to address this issue, simulations were done with continuous
star formation scenarios with different SFRs and durations as
presented in Figs. A.10, B.5 and B.10. It was found that a SFR of
several tens of M$_{\sun}$~yr$^{-1}$ is required to reproduce the
fluxes found in observational data.  Simulations with a duration of
100 Myr show that after about 10-20 Myr there is no further evolution
in the IR fluxes and colors.  However, our data cannot exclude longer
burst durations, but there are considerations that would make such
scenarios unlikely. First of all, long periods of star formation at
high rates would require vast amounts of gas. Secondly, these high
star formation rates also produce a lot of stellar feedback (stellar
winds, SNe) that makes it increasingly difficult to supply gas to form
new stars.

Because of these reasons, we consider the case of a high SFR and a
short duration (100 M$_{\sun}$~yr$^{-1}$ and 10 Myr, leading to the
formation of $10^9$ M$_{\sun}$ ~of stars, and making it comparable to
the other simulations).  The results of this simulation are compared
to DEFAULT in Fig.  B.5. This figure shows some
remarkable differences between the two situations. Whereas the
instantaneous burst starts at a maximum and decays rapidly after about
3 Myr, the continuous case starts at much lower values and takes quite
some time to rise. After about 5 Myr the two scenarios cross and the
continuous case starts to dominate. After about 10 Myr it reaches its
maximum, but it does not decay, since stars keep forming at a constant
rate.

The significant difference between the continuous and the
instantaneous star formation is that objects with instantaneous star
formation are only luminous enough in the first 5 to 10 Myr, which
decreases their detection rate. The continuous starbursts remain
luminous as long as there is inflow of gas to feed the star formation.

\subsection{Stage 3 - Stellar dust}
\label{sec:other-simulations}

The torus dust alone is not able to explain all the IR 
continuum properties and therefore in the second set of
simulations addresses the properties of the stellar dust
component. The lack of emission at short wavelengths can have two
reasons. A very simple explanation is that the geometry we are
currently using does not provide enough hot dust and more dust
must be added in the center of the dust distribution. Another
possibility is that the optical properties of the large-scale dust
are inappropriate and that the stellar dust modelling requires
more realistic optical properties.

\subsubsection{Smaller radius}
\label{sec:smaller-radius}
One of the ways to create more hot dust is to decrease the size of the
total dust distribution. This way all the dust will be closer to the
central source and therefore the overall temperature will be
higher. In order to investigate this possibility, two simulations were
carried out, with $r_0=50$ pc and $r_0=30$ pc. The rest of the
parameters were the same as for DEFAULT. As can be seen in
Fig. B.6, decreasing $r_0$ does improve the
short wavelength results. However, it also affects the longer
wavelengths and although the fit in the 10$\mu$m/100$\mu$m CLD is
better, the slope is not correct. A positive effect is that is
introduces scatter in the vertical direction of the CLD, which could
not be explained well by the previous simulations.  Overall one can
conclude that although changing the radius of the dust geometry does
improve the results, it does not provide the hot dust
contribution that we are looking for.

\subsubsection{Grey stellar dust}
\label{sec:extra-dust-component} Another possible solution is to
add additional stellar dust close to the starburst. Maybe our
assumption that the dust close to the central region is completely
destroyed or blown away is not correct. Therefore, we introduce an
extra (Gaussian) dust shell with a scale size of 10 pc (see also
Fig. \ref{fig:dust-T}). Two simulations are performed, one where the
total column of both the stellar and the torus dust is scaled to an
optical depth of 10 and one where the dust density of the stellar
component is first multiplied by a factor of 10 and then the total
column is scaled to an optical depth of 10.  Again, the rest of the
parameters are the same as for DEFAULT. Fig.
B.7 shows that introducing this second
component does increase the 10$\mu$m flux, without affecting the
60$\mu$m and 100$\mu$m fluxes. However, the 20$\mu$m flux is still
influenced, and the 10$\mu$m flux cannot be increased further, without
compromising the 20 $\mu$m results.  These simulations also introduce
scatter perpendicular to the evolutionary tracks.

Both attempts to solve the 10$\mu$m problem by introducing extra
hot dust, do not provide the results that are desired.  This
suggests that the optical properties of the dust must be
modified.

\subsubsection{Non-grey stellar dust}
\label{sec:optic-dust-prop}

Since the stellar dust component must have non-grey optical
properties, it it will be composed from graphite, silicates and
amorphous carbon \citep[as described in][]{1984ApJ...285...89D}.
In these computations, the SB99 output spectrum is first processed
by the non-grey stellar dust and the emergent radiation is used to
irradiate the (grey) torus dust distribution.

As can be seen in Fig. \ref{fig:final-results} (see also Fig.
A.11 and Fig. B.8 for
the full results and Fig. \ref{fig:final-SED} and Fig.
A.12 for SEDs), the addition of non-grey
stellar dust finally produces the right amount of 10$\mu$m
radiation, without influencing the flux at longer wavelengths. In
addition, although this second dust component has a large
influence on the 10$\mu$m radiation, {only a very low optical
depth is needed, which corresponds to a small amount of dust (see
Fig. B.9).  The final results are
created using an V-band optical depth for the second component
($\tau_2$) of only 0.1. In terms of mass this means that the
amount of dust in the second component is about 0.1\% of the total
dust mass. Observationally, similar values are found.  For
instance, \cite{1997A&A...325L..21K} found that in order to
properly fit the SEDs of Arp 244, NGC 6240, and Arp 220, they need
to add about 0.1\% of hot dust to the bulk of the (colder) dust.

Physically, the stellar dust component most likely represents the
left-overs of the dust, which was shocked by stellar ejecta and
blown away by the stellar radiation. This dust is much closer to
the stars than the torus dust, is hotter and experiences
disruptive events like sputtering and shattering. A simple grey
dust model does not give the right results, since for these
smaller grain sizes the wavelength dependence of the absorption
cross section is important
\citep[e.g.][]{2003PhDT........18S,2004ApJ...613..986P}.  The
torus dust is the ring or shell of dust surrounding the starburst
nucleus. It has higher column densities, larger grain sizes and is
further away, giving rise to lower dust temperatures. 
Therefore, the grey approximation is valid for this dust
component.

\begin{figure}
  \centering
  \includegraphics[angle=0,width=\hsize]{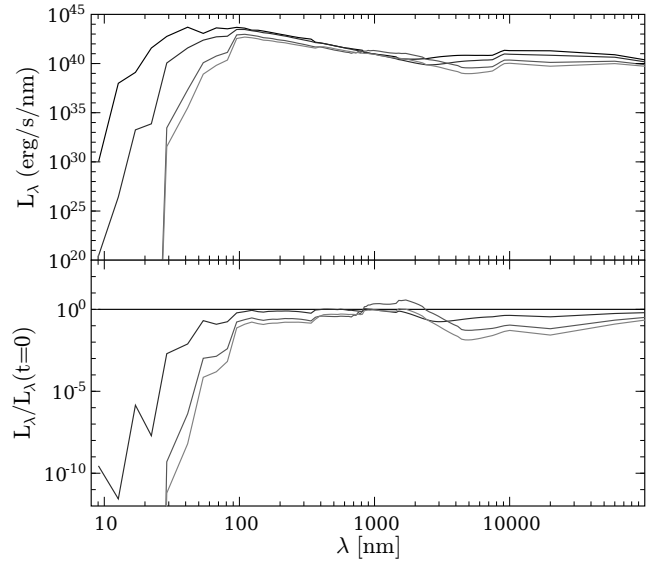}
  \caption{These panels show the evolution of the emergent spectrum of
  our final simulation. The top panel shows four spectra as a function
  of time ranging from t=0 (black) to t=15Myr (light grey) in steps of
  5Myr. The bottom panel shows the same spectra, but this time divided
  by the spectrum of t=0. All spectra were taken from the face-on
  results.}
  \label{fig:final-SED}
\end{figure}

\section{Conclusions}
Our simulation model is able to reproduce the IR  continuum
properties of starburst galaxies, by using two dust components: a
large scale component containing the bulk of the mass and a less
massive component close to the starburst.  The long wavelength
radiation is influenced by macro-physics like the star formation
activity and the large scale dust distribution. The short
wavelength emission at 10 $\mu$m comes from hot dust and is
influenced by the stellar dust and micro-physics like the optical
properties and optical depth of the dust.

Not all parameters have a profound effect on the results. The stellar
dust optical depth ($\tau_2$) and the parameters controlling the star
formation (the total stellar mass ${\rm M_{SB}}$ and the IMF slope
$\alpha$) have a large effect on the final results, whereas the
influence of the large scale dust geometry ($\tau$ and $\eta$) on the
final IR properties is smaller.\\
\\
{\bf Star formation region}\\
Because of their large influence on the results, the star formation
parameters could be determined well.
Increasing the stellar mass (${\rm M_{SB}}$) has two effects. First of
all, the shape of the spectrum changes, the short wavelength fluxes
increase as compared to the long wavelengths. A second effect is an
increase in ${\rm L_{IR}}$, which changes a factor of 10 when changing
the mass by a factor of 10.
The effects of varying the IMF slope ($\alpha$) are similar to those of the
stellar mass, but are less pronounced when decreasing $\alpha$ than when
increasing it. Making the IMF slope more shallow increases the luminosity
by a factor of 4.0, whereas steepening the slope decreases the
luminosity by a factor of 7.9.
There is a degeneracy between ${\rm M_{SB}}$ and $\alpha$. A less
massive starburst with a shallower IMF will produce roughly the
 same amount of OB stars as a more massive starburst with a
steeper IMF and therefore the same amount of massive stars and
therefore the same irradiating UV flux. Assuming the stars are
formed according to a Salpeter IMF ($\Psi(M) \propto M^{-2.35}$),
the star formation region should produce $10^9$~M$_{\sun}$ ~of
stars (either in one instantaneous burst or in a continuous
process) in order to
produce enough IR radiation.  \\ \\
{\bf Stellar dust}\\
For the stellar dust component, the grey approximation of the optical
dust properties is not valid. A more realistic dust model, including
graphite, silicates and amorphous carbon, is necessary to produce the
right results.
In addition, the V-band optical depth ($\tau_2$) is found to
be an important factor.  Using a $\tau_2$ of 10 (or 1)
overproduces the amount of 10$\mu$m radiation by almost a
factor of 4 (or 2.5). We require an
optical depth of 0.1 to get values that agree with observations.\\ \\
{\bf Torus dust}\\
The influence of the large scale dust geometry on the final IR
properties was far less and therefore these parameters could not be
determined to great precision.
Increasing the dust column density ($\tau$) has two effects. Like for
${\rm M_{SB}}$, the short wavelengths are enhanced compared to the
long wavelengths and the IR luminosity increases. These effects are,
however, much smaller: the luminosity changes by a factor of 2.0 when
going from $\tau$=1 to 10 and by a factor of 3.2 when increasing it
further to 100.  A second effect is that the inclination dependence
increases with increasing optical depth. The difference in the edge-on
and face-on values of ${\rm L_{IR}}$ changes by a factor of 1.6 for
$\tau=1$ to a factor of 16 for $\tau=100$.
The best fit to the data was obtained with $\tau=10$, but the other
simulations were also reasonable.
 The closing angle did not seem to have an optimum value at
all. Varying it only affects the inclination dependence of the
results. The variation of ${\rm L_{IR}}$ for a flat disk-like
structure ($\eta=0.3$) is about a factor of 6.3, compared to a factor
of 2.0 for a more shell-like dust geometry with a closing angle of 0.7.
Even though the inclination effects are reduced for higher values,
the edge-on results still do not match the data.\\

Considering the values determined for the parameters
investigated, it seems that, although observationally starburst
galaxies appear to have a very violent nature, the star formation
environment does not need to be as ``exotic'' as one might expect.
The only exceptional parameter needed to explain the high IR
output of starburst galaxies is the large amount of massive stars
(high ${\rm M_{SB}}$ or SFR), but there is no need for an adjusted
IMF to increase the number of heavy stars.  The torus dust
surrounding the stars is no exception to this trend. Both the size
(100 pc) and the V-band optical depth of the torus dust  are
moderate (10, comparable to values found for photon dominated
regions). Furthermore, most of the dust has a low temperature
($\sim 75$K at 100pc from the center),  while only a very
small amount of hot dust
($\sim 400$K at 10pc) is needed.\\

In all simulations, including the final ones, the data were only
fit well by the face-on evolutionary tracks. All the completely
edge-on results were a poor fit. This effect is stronger in the
short wavelength results than in the long wavelength result. This
indicates that inclination dependence is mostly caused by the
obscuration of the hot dust in the center by the outer dust
distribution.  This can have two implications. On the one hand, it
could be that the inclination dependence resulting from our model
is too large, and that extra parameters  are needed to
address this problem. A likely parameter is the clumpiness of the
dust. A clumpy medium with the same average density (i.e. the same
mass) would have a lower apparent optical depth then a smooth
medium \citep[e.g.][]{1984ApJ...287..228N,2005Ap&SS.295..319C},
which makes it easier for the short wavelength radiation to travel
in the edge-on direction.  On the other hand, if the predictions
of our model are correct, the implication is that there is a
significant number of starburst ULIRGs, which are currently not
classified as such based on their IRAS colors.

A specific shortcoming of our models is that, although the IR
properties are well explained, almost all parameters move the
evolutionary tracks more or less along the same line and in the
direction of the evolution. The result is that not all parameters
of the physical environment in a given starburst galaxy can be
uniquely inferred from observations, using this model. The UV
input can be inferred from the total IR output, but this does not
constrain the IMF or the SFR. Similarly, the optical properties of
the dust can be determined, but the geometry is hard to infer.
To address these problems, more information than just the IR
continuum is needed and therefore we intend to extend the current
model. First of all, the IR part of the code will be modified to
include specific spectral characteristics (e.g. PAHs and high
ionization lines). Also the molecular environment that surrounds
the current dust region will be added to the model. The chemistry
of such a region will give a better handle on the radiation field,
as well as the densities and temperatures of the gas
\citep[e.g.][]{1999RvMP...71..173H, 2005A&A...436..397M,
2002A&A...381..783A, 2004A&A...419..897U, 2004ApJS..152...63G,2005ApJ...629..767O,
2006ApJ...640L.135G,2006A&A...000..000B} . Also other wavelengths
will be studied, since optical and UV data will give more
information about the star forming region, whereas (sub)millimeter
and radio observations will reveal more about the outer dust
regions and the molecular environment.

\begin{acknowledgements}
AFL would like to thank Rowin Meijerink and Kees Dullemond for
their extensive support on setting up and using RADMC and RADICAL and
Claus Leitherer for answering all questions about STARBURST99.
\end{acknowledgements}

\bibliographystyle{aa}
\bibliography{5778}
\end{document}